\documentclass[12pt]{article}
\usepackage[dvips]{graphicx}
\usepackage{epsfig,cite}
\usepackage {amsmath}
\usepackage{color}
\usepackage{amssymb}
\usepackage{slashed}
\include{epsf}
\newcount\timecount
\newcount\hours \newcount\minutes  \newcount\temp \newcount\pmhours
\hours = \time
\divide\hours by 60
\temp = \hours
\multiply\temp by 60
\minutes = \time
\advance\minutes by -\temp
\def\hour{\the\hours}
\def\minute{\ifnum\minutes<10 0\the\minutes
            \else\the\minutes\fi}
\def\clock{
\ifnum\hours=0 12:\minute\ AM
\else\ifnum\hours<12 \hour:\minute\ AM
      \else\ifnum\hours=12 12:\minute\ PM
            \else\ifnum\hours>12
                 \pmhours=\hours
                 \advance\pmhours by -12
                 \the\pmhours:\minute\ PM
                 \fi
            \fi
      \fi
\fi
}

\def\monthname{\relax\ifcase\month 0/\or January\or February\or
   March\or April\or May\or June\or July\or August\or September\or
   October\or November\or December\else\number\month/\fi}

\def\bold#1{\setbox0=\hbox{$#1$}%
     \kern-.025em\copy0\kern-\wd0
     \kern.05em\copy0\kern-\wd0
     \kern-.025em\raise.0433em\box0 }

\def\beq{\begin{equation}}
\def\eeq{\end{equation}}

\def\ga{\mathrel{\raise.3ex\hbox{$>$\kern-.75em\lower1ex\hbox{$\sim$}}}}
\def\la{\mathrel{\raise.3ex\hbox{$<$\kern-.75em\lower1ex\hbox{$\sim$}}}}
\def\gev{{\rm \, Ge\kern-0.125em V}}
\def\tev{{\rm \, Te\kern-0.125em V}}
\def\gyr{{\rm \, G\kern-0.125em yr}}

\def\gappeq{\mathrel{\rlap {\raise.5ex\hbox{$>$}}
{\lower.5ex\hbox{$\sim$}}}}
\def\lappeq{\mathrel{\rlap{\raise.5ex\hbox{$<$}}
{\lower.5ex\hbox{$\sim$}}}}
\def\Toprel#1\over#2{\mathrel{\mathop{#2}\limits^{#1}}}

\def\m12{m_{1\!/2}}

\def\bea{\begin{eqnarray}}
\def\eea{\end{eqnarray}}

\def\beqar{\begin{eqnarray}}
\def\eeqar{\end{eqnarray}}

\begin{document}

\begin{titlepage}
\pagestyle{empty}
\baselineskip=21pt
\rightline{}
\vskip 0.8in
\begin{center}
{\Large\bf{ Probing the Anomalous FCNC Interactions in Top-Higgs Final State and Charge Ratio Approach}}

\end{center}
\begin{center}
\vskip 0.4in
{\bf Sara Khatibi and Mojtaba Mohammadi Najafabadi }
\vskip 0.1in
{\it  School of Particles and Accelerators, \\
Institute for Research in Fundamental Sciences (IPM) \\
P.O. Box 19395-5531, Tehran, Iran}\\
\vspace{2cm}
 \textbf{Abstract}\\
 \end{center}
\baselineskip=18pt \noindent
{
We study the anomalous production of a single top quark in association with
a Higgs boson at the LHC originating from flavor changing neutral current (FCNC)
interactions in $tqg$ and $tqH$ vertices.
We derive the discovery potentials and $68\%$ C.L. upper limits
considering leptonic decay of the top quark and
the Higgs boson decay into a $b\bar{b}$ pair
with 10 fb$^{-1}$ integrated luminosity of data in proton-proton collisions at the
center-of-mass energy of 14 TeV.
We propose a charge ratio for the lepton in top quark decay
in terms of lepton $p_{T}$ and $\eta$ as a strong tool to observe the signal.
In particular, we show that the charge ratio increases significantly at large $p_{T}$ of the charged lepton.
While the main background from $t\bar{t}$ is nearly charge symmetric and $W+jets$ background has much
smaller charge ratio with respect to the signal. We show that this feature can also be used in the probe of 
anomalous single top production with a $Z-$boson or a photon which are under the attention of the experimental 
collaborations.}

\vfill
\leftline{PACS numbers: 14.65.Ha,12.60.-i,12.38.Bx,14.80.Bn}
\end{titlepage}
\baselineskip=18pt


\section{Introduction}

\begin{figure}
\centering
  \includegraphics[width=7cm,height=5cm]{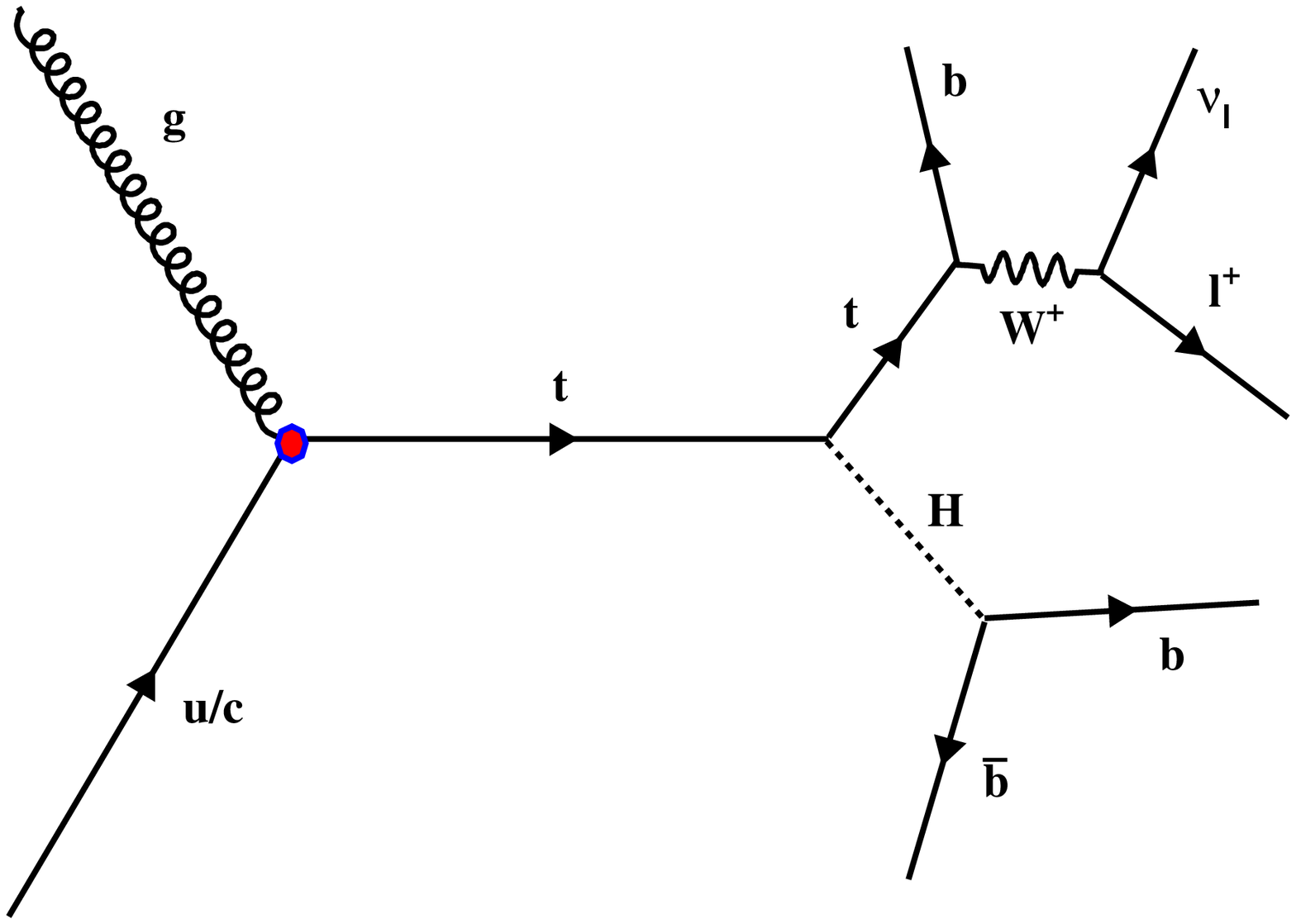}
  \includegraphics[width=7cm,height=5cm]{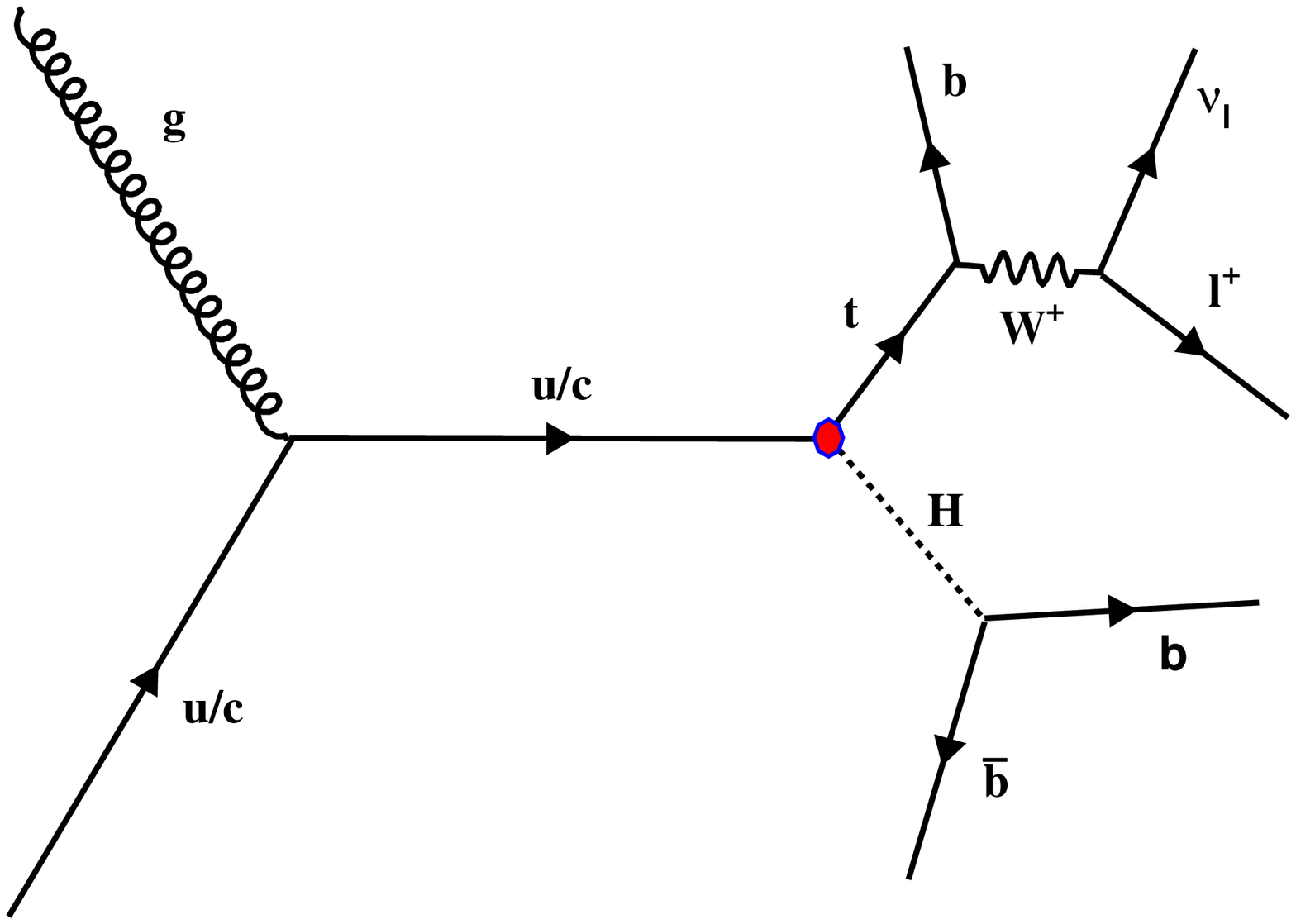}
  \caption{The representative Feynman diagram for production of a top quark in association with a Higgs boson including the
decay chain with leptonic top quark decay and Higgs decay into a $b\bar{b}$ pair.}\label{feynman}
\end{figure}

In view of the top quark large mass, it is a unique place to probe the dynamics that breaks the
electroweak gauge symmetry.
Several properties of the top quark have been measured and studied using the
data collected with the LHC experiments at the center-of-mass energies of 7 and 8 TeV
as well as the Tevatron experiments.
The top quark interacts with other Standard Model (SM) particles via gauge and Yukawa
interactions.
So far, many remarkable results have come out of the LHC and Tevatron experiments
including the top quark interactions in
both electroweak and strong sectors.
It is worthy to mention that both the ATLAS and the CMS experiments
have measured several properties of the top quark with high precision \cite{frank},\cite{werner}. In particular,
the cross section for single top production has been measured with a precision of less than
$15\%$ \cite{st} and the present measurement of the top pair rate is better than $10\%$ \cite{tt}.
Undoubtedly,  it is expected that the top quark
properties will be measured with more precision using more amount of data and
in the next phase of the LHC with collisions at 13 or 14 TeV.\\
Flavor Changing Neutral Current (FCNC) couplings are strongly suppressed in top
sector at tree level in the SM framework by Glashow-Iliopoulos-Maiani (GIM)
mechanism \cite{Glashow:1970gm}. While the FCNC processes involving the top quark can appear in
models beyond the SM. In particular, significant FCNC
couplings of top quark with an up or charm quark
and a gluon are predicted in several new physics models beyond the SM \cite{Tait:2000sh},\cite{AguilarSaavedra:2004wm},\cite{Liu:2004qw}, \cite{Bejar:2008ub},\cite{Cao:2008vk} \cite{GonzalezSprinberg:2006am}, \cite{AguilarSaavedra:2002ns}.
The anomalous FCNC couplings for top with an up-type quark (u,c) and a gluon
can be described in a model independent effective Lagrangian way according to the following \cite{Malkawi:1995dm},\cite{Hosch:1997gz}:
\begin{eqnarray}\label{lag}
\mathcal{L}=\sqrt{2}g_{s}\sum_{q=u,c} \frac{\kappa_{tqg}}{\Lambda} \bar{t} \sigma^{\mu\nu} T_{a}
(f_{q}^{L} P_{L}+f_{q}^{R} P_{R})q G_{\mu\nu}^{a}+h.c.
\end{eqnarray}
Here $P_{L}$ and $P_{R}$ are chirality projection operators.
In Eq.\ref{lag}, $\Lambda$ is the energy scale which new physics appears and $\kappa_{tqg}$ are real
dimensionless parameters thus $\frac{\kappa_{tqg}}{\Lambda}$ are the strength of the
couplings. The parameters $f_{q}^{L}$ and $f_{q}^{R}$ are chiral parameters with the normalization of $
|f_{q}^{L}|^{2}+|f_{q}^{R}|^{2}=1$.\\
There are many analyzes in search for the
anomalous $tqg$ and other anomalous couplings related to
the top interaction in the literature \cite{Gao:2011fx}, \cite{Wang:2012gp},\cite{Agram:2013koa},\cite{Kidonakis:2003sc},\cite{Etesami:2010ig},\cite{Zhang:2008yn}.
The CDF and D0 experiments at the Tevatron have searched for these FCNC couplings \cite{Aaltonen:2008qr},\cite{Abazov:2010qk}.
The $95\%$ confidence level limits on the anomalous FCNC couplings have been found to be:
\begin{eqnarray}\label{atlasbound}
\frac{\kappa_{tug}}{\Lambda}<0.013~\text{TeV}^{-1}~,~~
\frac{\kappa_{tcg}}{\Lambda}<0.057~\text{TeV}^{-1}
\end{eqnarray}
Recently, the ATLAS experiment set $95\%$ C.L. upper limits on the strong FCNC couplings using
$14.2$ fb$^{-1}$ of 8 TeV data. In the ATLAS search for the FCNC events in $tqg$ vertex,
the production of a single top quark with or without another light quark or gluon are  considered \cite{atlas}.
The extracted limits are the most stringent limits on these couplings:
\begin{eqnarray}\label{aaa}
\frac{\kappa_{tug}}{\Lambda}<5.1\times 10^{-3}~\text{TeV}^{-1}~,~~
\frac{\kappa_{tcg}}{\Lambda}<1.1\times 10^{-2}~\text{TeV}^{-1}
\end{eqnarray}
The FCNC anomalous interaction $tqg$ can lead to production of a top quark in association with a $Z-$boson.
In \cite{cmstz}, a search for the top quark anomalous couplings has been performed through the search for
the final state of a single top quark in association with a $Z-$boson at the LHC with the CMS detector.
This search has been performed using 5 fb$^{-1}$ of proton-proton collisions at 7 TeV data.
 The $95\%$ C.L. observed upper limits on the anomalous couplings of the effective model are found to be:
\begin{eqnarray}
\frac{\kappa_{tug}}{\Lambda}<0.1~\text{TeV}^{-1}~,~~
\frac{\kappa_{tcg}}{\Lambda}<0.35~\text{TeV}^{-1}
\end{eqnarray}
Lower cross section of this process and smaller amount of data is the reason that these bounds are
looser than the bounds with respect to the bounds indicated in Eq.\ref{atlasbound}, Eq.\ref{aaa}. However,
performing such an analysis is necessary to check the consistency of all results in searches for FCNC.
There is another detailed study for the anomalous interactions of $tqg$ using the $tZ$ channel at the
LHC with 20 fb$^{-1}$ of 8 TeV collisions in \cite{Agram:2013koa}.
The $3\sigma$ discovery ranges obtained in this study are as follows:
\begin{eqnarray}
\frac{\kappa_{tug}}{\Lambda} > 0.09~\text{TeV}^{-1}~,~~
\frac{\kappa_{tcg}}{\Lambda} > 0.31~\text{TeV}^{-1}
\end{eqnarray}
The discovery of a new Higgs-like particle with a mass of around 125 GeV by the
ATLAS and CMS experiments at the LHC \cite{Aad:2012tfa}, \cite{Chatrchyan:2012ufa} has opened a new window in searches
for different properties of SM particles. In particular, because of the large coupling of
the Higgs boson with top quark, the top quark properties
could be studied in channels where a Higgs boson is also present.
In this work, we perform a search for anomalous top interaction of
$tqg$ by studying a signature consisting of a Higgs boson and a single top quark.
We perform the analysis for 10 and 100 fb$^{-1}$ of the LHC
proton-proton collisions at the center-of-mass energy of
14 TeV. We investigate the final state of three b-jets where the top quark
decays to a charged lepton (muon or electron), neutrino and a b-quark and the
Higgs boson decays into a $b\bar{b}$ pair.
The representative Feynman diagram of the signal process including the
decay chain is shown in Fig.\ref{feynman} (left).
In the final state we expect only one charged lepton, missing energy and three $b$-tagged jets.
We find the parameter regions where the LHC may be able to observe the signal, otherwise
upper limits are set on the anomalous couplings. 
The real data of the LHC could be used in search for the anomalous $tqg$ couplings in this channel
 since it provides really reasonable results in comparison
with the already obtained results from other channels even with a simple set of cuts.
In order to improve the sensitivity to the
$tqg$ anomalous couplings, the $tH$ channel results can be combined with
both the FCNC single top quark and top pair production modes. \\
It is remarkable that for our favorite signal the radiation of a Higgs boson does not change the spin direction of the top quark.
Therefore, if the anomalous interactions are quite left-handed ($f_{q}^{L} = 1,f_{q}^{R}=0$)
or right-handed ($f_{q}^{L} = 0,f_{q}^{R}=1$), the top quark is produced with the spin direction
parallel to the incident quark momentum direction for the left-handed case and opposite to the
incident quark momentum for right-handed case. The chirality information is transferred to the
decay products of the top quark, accordingly by a careful study of the charged lepton angular
distribution, the type of interaction (left-handed or right-handed couplings) could be determined.
Among the channels by which we can probe the anomalous $tqg$ couplings, the direct top production
\cite{Gao:2011fx} and top plus Higgs channel ($u(c)+g\rightarrow t+H$) provide
the possibility to determine the chirality nature of these couplings.\\
It is interesting to note that in addition to the effective FCNC Lagrangian in the vertex of $tqg$, introduced in
Eq.1, the anomalous FCNC interaction in the $tqH$ vertex leads to production of a single top 
quark in association with a Higgs boson as well. This is illustrated by a Feynman diagram
in the right-side of Figure \ref{feynman} where flavor changing interaction of the
top quark and light quark invloves a Higgs boson.
The anomalous FCNC interaction $tqH$ can be 
parametrized as the following \cite{AguilarSaavedra:2004wm}:
\begin{eqnarray}
\mathcal{L}=\frac{g}{2\sqrt{2}}\sum_{q=u,c}g_{tqH}\bar{q}(g^{v}_{tqH}+g^{a}_{tqH}\gamma_{5})tH+h.c.
\end{eqnarray}
where the real coefficient $g_{tqH}$ ( with $q = u, c$) denotes the strength of the anomalous coupling. 
The coefficients $g^{v}_{tqH}, g^{a}_{tqH}$ are in general
complex numbers with the normalization $|g^{v}_{tqH}|^{2}+|g^{a}_{tqH}|^{2}=1$.
The $95\%$ C.L. upper bounds on the FCNC $tqH$ couplings deriveded from the low energy
experiments with the Higgs boson mass in the interval of 115 to 170 GeV are \cite{tqh1},\cite{tqh2}:
\begin{eqnarray}
g_{tuH} < 0.363-0.393~,~g_{tcH} < 0.270-0.319
\end{eqnarray}
In \cite{Wang:2012gp} the anomalous production of a single top quark with a Higgs boson
via the FCNC interaction of $tqH$ has been studied at the LHC including complete QCD
next-to-leading order corrections. The $3\sigma$ exclusion upper limits on the anomalous couplings with the  Higgs boson mass
of 125 GeV based on 10 fb$^{-1}$ of the inegrated luminosity have been found to be:
\begin{eqnarray}
g_{tuH} < 0.121~,~ g_{tcH} < 0.233
\end{eqnarray}
It is noteable that both anomalous couplings $tqg$ and $tqH$ are arising from dimension-six operators. Therefore, it makes sense 
to consider both anomalous interactions together. The Feynman diagrams depicted in Figure \ref{feynman} can be studied 
simultaneously that leads to an interference term. In this paper, we sutdy the single top plus a Higgs boson final state
once in the presence of only $tqg$ couplings and once in the presence both $tqg$ and $tqH$ anomalous interactions.

The organization of this paper is as follows. Next section is devoted to events
simulation for signal (left diagram of Figure \ref{feynman}) and backgrounds and analysis. In section 3, we obtain the discovery
potential and $68\%$ C.L. upper limits on the anomalous couplings from this channel and discuss the results. 
Section 4 presents a simultaneous probe of $tqg$ and $tqH$.
In section 5, we will discuss a way based on the leptonic charge ratio
to discriminate between signal and backgrounds and to distinguish between $tug$ and
$tcg$ couplings. In particular, we look at the charge ratio as a function of $p_{T}$ and $\eta$ of the charged lepton for signal and background
processes. Finally, conclusions are presented in section 6.

\section{Event Simulation and Selection}

\begin{figure}
\centering
\includegraphics[width=7cm,height=6cm]{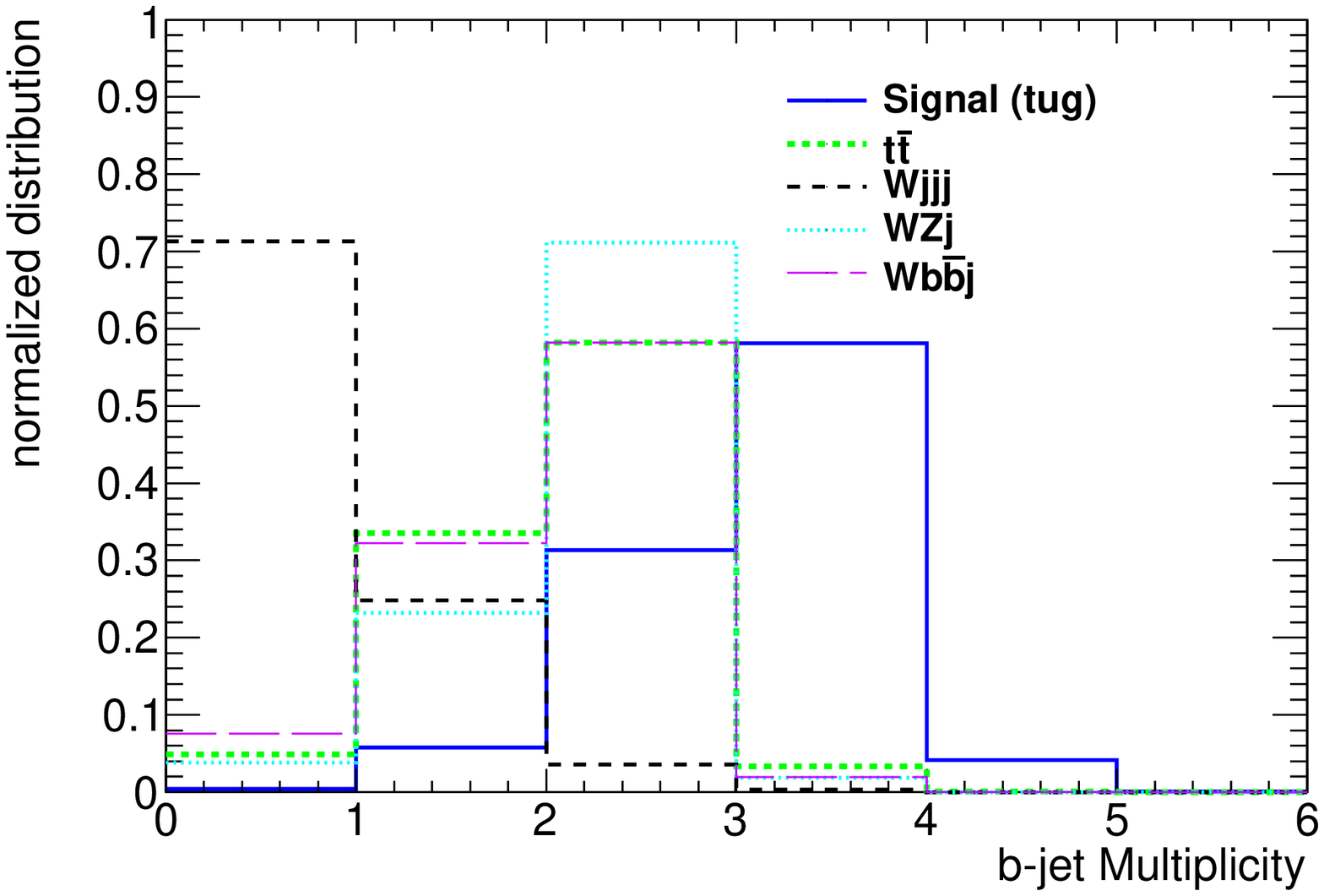}
\includegraphics[width=7cm,height=6cm]{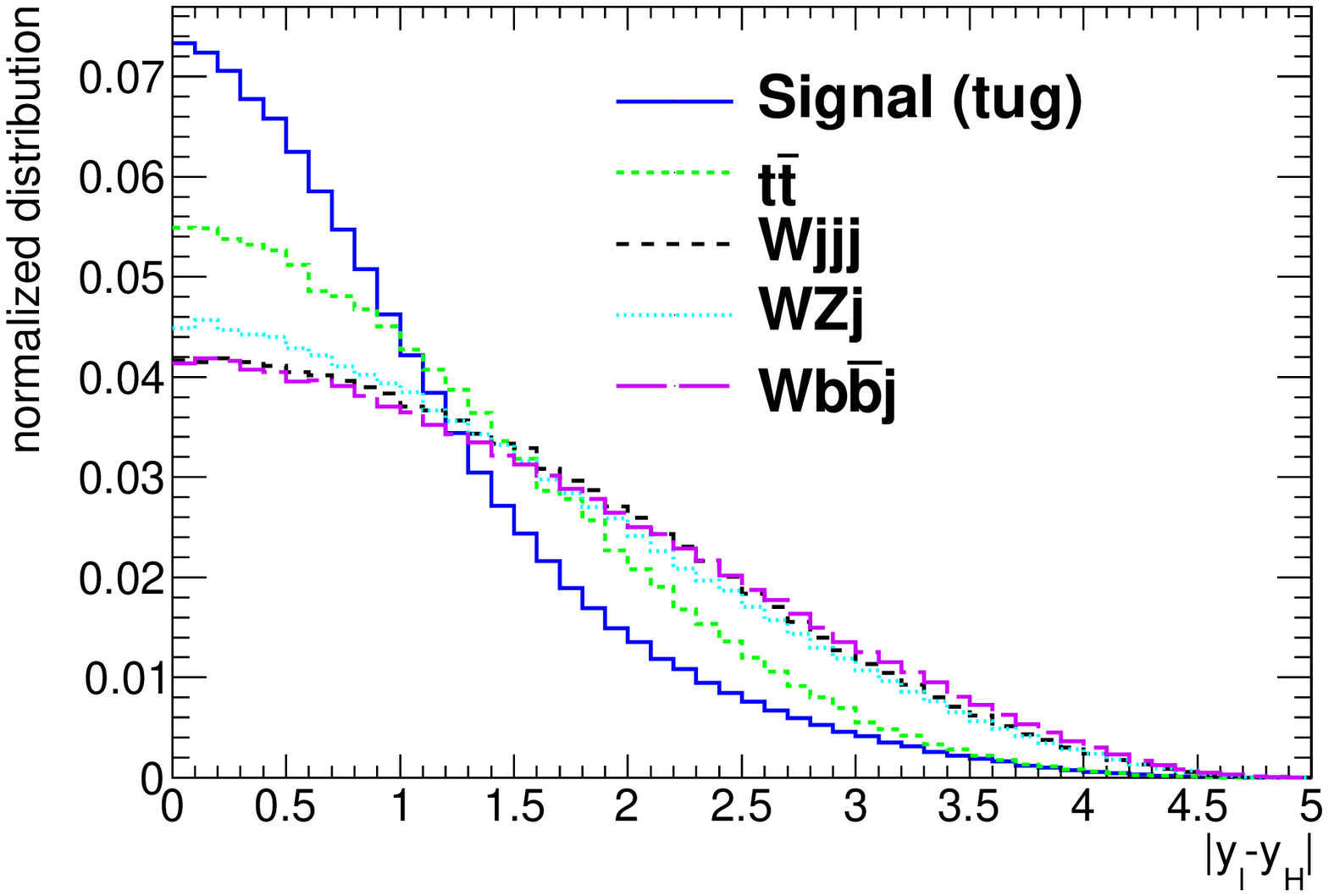}
\caption{b-jet multiplicity distribution for signal and backgrounds (left) and
the reconstructed distribution of $|y_{l}-y_{H}|$ for signal and different backgrounds. The distributions are
normalized to one.}\label{variables1}
\end{figure}

In this section, we define the signal and backgrounds processes and
describe the simulation method, the event selection and reconstruction
of the final state.
The process of signal is taken as the single top plus a Higgs boson followed
by the leptonic top quark decay and the Higgs boson decay into a $b\bar{b}$
pair. The Feynman diagram of production and decay chain is presented in Figure \ref{feynman}.
The main background processes are $Wb\bar{b}j$, $Wjjj$, $WZj$, and $t\bar{t}$.
For both signal and the background processes, the {\sc MadGraph} 5 package \cite{Alwall:2011uj}
has been used to generate the hard scattering matrix elements with the {\sc cteq6} \cite{Pumplin:2002vw} as 
parton distribution function. The parton level events are passed through {\sc Pythia} 8 \cite{Sjostrand:2007gs}
for showering. The jet reconstruction is then
performed by {\sc  Fastjet} package \cite{Cacciari:2005hq} using an anti-$k_{t}$ algorithm
with the cone size of $R = 0.5$ \cite{Cacciari:2008gp}. Where $R = \sqrt{(\Delta\eta)^{2}+(\Delta\phi)^{2}}$, with
$\eta=-\ln\tan(\theta/2)$. The parameters $\eta$ and $\phi$ are the polar and azimuthal
angles w.r.t the $z-$axis.
In this analysis, we focused on the LHC run with the center-of-mass energy of $\sqrt{s}=14~TeV$
for the integrated luminosities of 10 fb$^{-1}$ and 100 fb$^{-1}$.
In order to simulate the signal events, the effective Lagrangian of Eq.\ref{lag}
has been implemented within the {\sc  Feynrules} package \cite{Christensen:2008py},\cite{Duhr:2011se} then
 imported the model to a UFO module \cite{Degrande:2011ua}
and then inserted to the {\sc MadGraph} 5. The cross sections has been found to be consistent
with  {\sc CompHEP} package \cite{Boos:2004kh},\cite{Pukhov:1999gg}. In this analysis, we only concentrate 
on the case that $f_{q}^{L}=f_{q}^{R}=1$.
The signal is generated with top quark decay leptonically (muon and electron) and the
Higgs boson decaying into $b\bar{b}$. 
The $t\bar{t}$ background is generated in semi-leptonic decay mode. The $Wb\bar{b}j$, $Wjjj$, $WZj$
are generated with again leptonic decay of the $W-$boson and for the latter one the $Z-$boson decays into a $b\bar{b}$.
To simulate b-tagging, a b-tagging efficiency of $60\%$ is chosen for b-jets
and a mis-tagging rate of $10\%$ for other quarks.
The effects of detector resolution are simulated through Gaussian
energy smearing which is applied to jets and leptons with a standard deviation parameterized
according to the following:
\begin{eqnarray}
\frac{\sigma(E)}{{E}}=\frac{{a}}{\sqrt{{E(\text{GeV})}}}\oplus b
\end{eqnarray}
where $\sigma(E)$ indicates the energy resolution at the energy value of $E$, the symbol $\oplus$
represents a quadrature sum, and the energies are measured in GeV.
For resolutions of jets (leptons) we take the values of ATLAS detector \cite{Aad:2009wy},
$a=0.5(0.1)$  and $b=0.03 (0.007)$.
It is notable that the electron and muon energy resolutions have different
dependencies on the electromagnetic calorimetry and the charged particle tracking.
Nevertheless, the uniform
values for electromagnetic calorimetry energy resolution is used for the final
state lepton. It is more conservative for the energies under consideration in the analysis
than the capabilities of tracking.
In order to trigger the events, every event is required to have at least one charged lepton passing
through the cuts on the rapidity and transverse momentum. The typical value for
charged lepton $p_{T}$ cut is 25 GeV within the pseudorapidity range of $|\eta|<2.5$.
The missing transverse energy is required to be larger than 25 GeV.
The jets are required to have  $p_{T}>25$ GeV with pseudorapidities to be $|\eta|<2.5$.
The angular distance between the charged lepton and jets and all jets have to be $\Delta R _{lj,jj}>0.4$.
The cross section of signal after the above preliminary cuts including the branching ratios are:
\begin{eqnarray}
\sigma(\kappa_{tug}/\Lambda)~\text{pb}= 5.60\times[\frac{\kappa_{tug}}{\Lambda}]^{2}~,
~\sigma(\kappa_{tcg}/\Lambda)~\text{pb} =1.05\times[\frac{\kappa_{tcg}}{\Lambda}]^{2}
\end{eqnarray}
where $\kappa_{tqg}/\Lambda$ is in TeV$^{-1}$. The process $Wjjj$
has the largest cross section which is 230.0 pb considering the cuts and branching ratios.
The $t\bar{t}$ cross section after
the cuts and taking into account the branching ratios is 34.35 pb. The cross sections of
$Wb\bar{b}j$ and $WZj$ processes are 2.33 pb and 0.138 pb, respectively.
In order to reconstruct the top quark and Higgs boson in the final state, first we require to
have only three b-tagged jets in each event.
The plot in the left side of Fig. \ref{variables1} shows the b-jet multiplicity
in signal and different backgrounds events. As can be seen the requirement of
only three b-tagged jets is useful to reduce the contribution of the backgrounds.
We specifically apply such a requirement to suppress the large contributions of background events
originating from $Wjjj$.
To reconstruct the top quark, the full momentum of the
neutrino is needed. The missing transverse energy is taken as the transverse component of the neutrino
momentum. The $z-$component of the neutrino momentum is obtained by using the $W-$boson mass constraint:
$(p_{l}+p_{T,\nu}+p_{z,\nu})^{2} = m_{W}^{2}$. In most cases, there are two solutions
for the $p_{z,\nu}$. As a result, the combination of the charged lepton and two neutrinos leads to two $W-$bosons
which are combined with the three b-tagged jets separately. Among the six combinations, the combination which
gives the closest mass to the top quark mass is selected. The other remaining two b-jets are combined to reconstruct the Higgs boson.

\begin{figure}
\centering
\includegraphics[width=7cm,height=6cm]{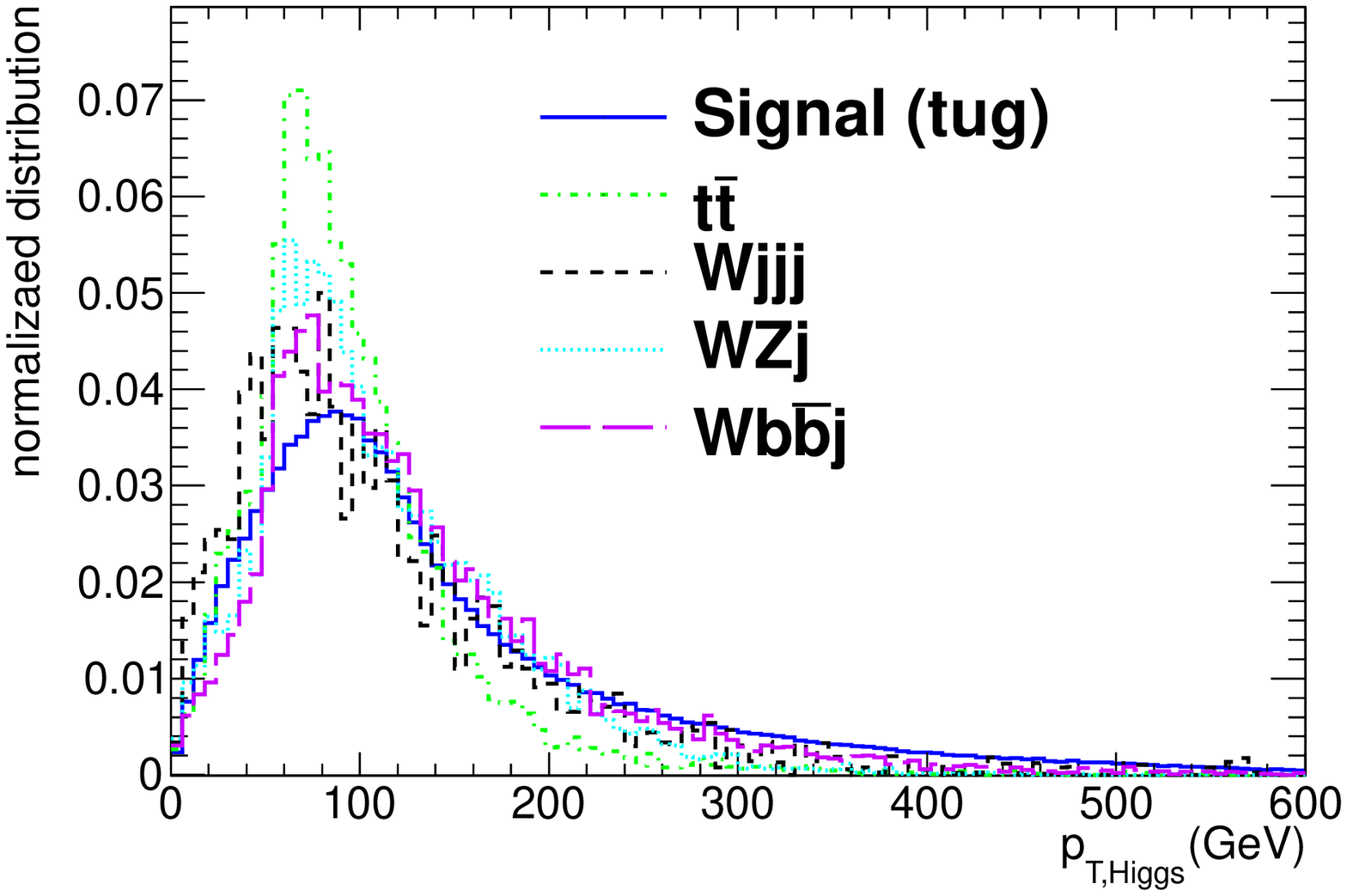}
\includegraphics[width=7cm,height=6cm]{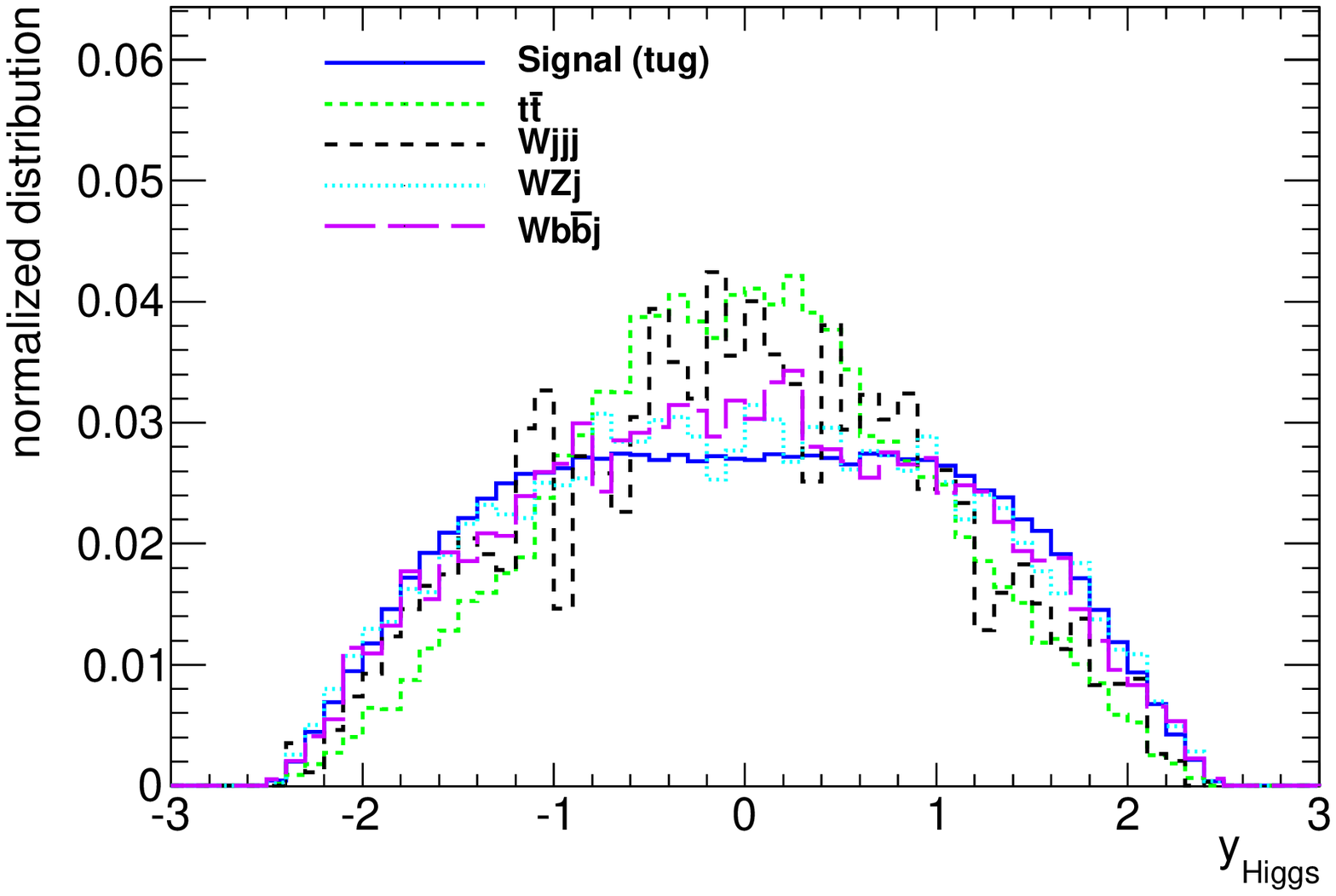}
\caption{The transverse momentum of the reconstructed Higgs boson (left) and the rapidity distribution of the Higgs boson (right)
for signal and backgrounds.}\label{variables2}
\end{figure}

\begin{figure}
\centering
\includegraphics[width=12cm,height=8cm]{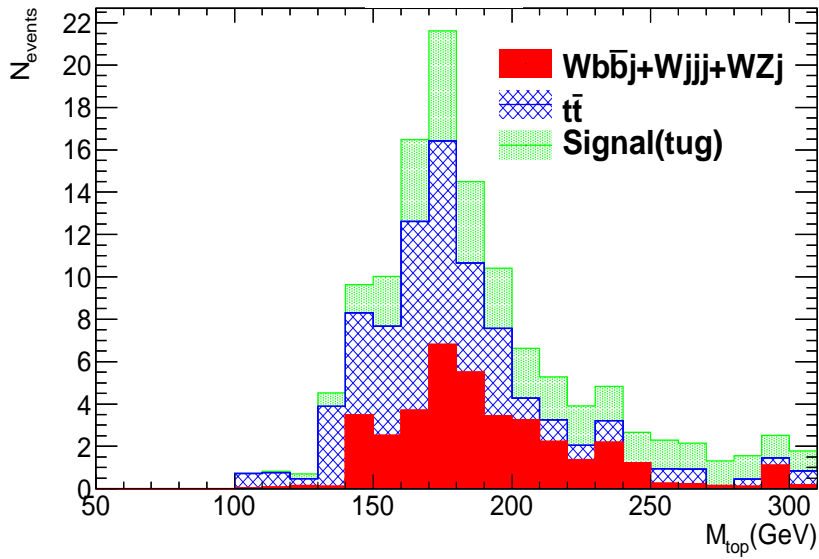}
\caption{{The reconstructed top mass distribution after all selection for 10 fb$^{-1}$
of LHC at 14 TeV center-of-mass energy for the signal with $\kappa_{tug}/\Lambda = 0.1$ TeV$^{-1}$ and backgrounds.}}\label{topmass}
\end{figure}

In order to suppress the backgrounds, we reject events with $|m_{H,rec}-125|>15$ GeV.
To reduce the contributions of the
backgrounds and enhance the signal contribution, we exploit some other kinematic distributions.
In the right panel of Fig.\ref{variables1}, the distribution of the difference between the
pseudorapidities of the charged lepton and the reconstructed Higgs boson ($|y_{l}-y_{H}|$)
is shown. The signal events prefer to reside mostly at around zero while the backgrounds in particular
the $W+jets$ events have a more spread distribution and is extended up to around 5. Therefore,
We require the events to satisfy $|y_{l}-y_{H}| < 1.2$ condition to reduce the $W+jets$
contributions.
We use two more kinematic variables to suppress the backgrounds.
In Fig.\ref{variables2}, the transverse momentum and rapidity
distributions of the reconstructed Higgs boson are depicted.
From the left panel of Fig.\ref{variables2}, we can see that in the $p_{T,Higgs}$ distributions
of $Wb\bar{b}j$, $Wjjj$, and $WZj$ the peaks are below 80 GeV while
for the signal the peak is around 90-100 GeV. Therefore, we require that
the transverse momentum of the Higgs boson to be greater than 100 GeV.
As it can be seen in the right panel of Fig.\ref{variables2},
for the signal process of $u+g\rightarrow t+H$, the Higgs bosons tend
to reside also in the forward and backward regions while the main backgrounds of $t\bar{t}$
and $W+jets$ are mostly central. Since the up quark on average carries larger
momentum with respect to gluon, the center-of-mass frame of the final state system is boosted
along the direction of the initial up quark.
We do not face with this situation for top pair events because the top pair
events are mostly coming from gluon-gluon fusions which are
symmetric. Only there is a small boost effect in top pair events due
to quark anti-quark annihilation. We choose the events
with $|y_{H}| > 0.8$. Because such an effect does not exist for
the signal process of $c+g\rightarrow t+H$, we do not apply this cut
for this process.
In Fig.\ref{topmass}, we show the reconstructed top quark mass after all cuts
for signal and backgrounds with 10 fb$^{-1}$ of integrated luminosity and with $\kappa_{tug}/\Lambda = 0.1$ TeV$^{-1}$.
It can be seen that top quark has been reconstructed well. 

\section{Results}

After applying all cuts which we explained in the previous section, we obtain the
following efficiencies for signal ($\frac{\kappa_{tug}}{\Lambda}= 0.1$ TeV$^{-1}$),
$t\bar{t}$,$Wb\bar{b}j$,$Wjjj$, and $WZj$ respectively:
$12\%$,$0.017\%$,$0.04\%$,
0.0023$\%$,0.071$\%$. For the $tcg$
signal the efficiency has been found to be $6\%$.
It should be mentioned that the $t\bar{t}$ process could also be considered as a source of top plus a Higgs boson. When one of the top
quarks radiates a Higgs boson, the final state consists of $t\bar{t}+H$.
Such events are unlikely to pass our selection. Because we require to have only one isolated
lepton as well as exactly three b-jets in the event which do not allow
such events to contribute to the signal.
We calculate the $3\sigma$ and $5\sigma$ discovery reaches of the LHC for the anomalous couplings $\frac{\kappa_{tug}}{\Lambda}$
and $\frac{\kappa_{tcg}}{\Lambda}$ after the event selection according to $S/\sqrt{B}$ formula.
The $3\sigma$ ($5\sigma$) values for 10 fb$^{-1}$ are summarized below:
\begin{eqnarray}
\frac{\kappa_{tug}}{\Lambda} \geq 0.069~(0.088)~{\text TeV}^{-1}~,~
\frac{\kappa_{tcg}}{\Lambda} \geq 0.26~(0.34)~{\text TeV}^{-1}
\end{eqnarray}
We see a better sensitivity to $\frac{\kappa_{tug}}{\Lambda}$ with respect to $\frac{\kappa_{tcg}}{\Lambda}$
which is because of the fact that the parton density function of the charm quark is suppressed
w.r.t the up quark.\\
The next-to-leading order QCD corrections to signal would improve
the results however the NLO corrections for our favorite signal
is not available. If we assume similar $k-$factor of 1.3
as direct top production ($g+u(c)\rightarrow t$) \cite{Gao:2011fx} and $g+u(c)\rightarrow t+Z$ \cite{Li:2011ek},
the results mentioned above will improve up to the order of $10\%$.
In case of finding no evidence for signal, upper limits can be set 
on the anomalous interaction parameters. To set the $68\%$ C.L. limits,
we use a simple $\chi^{2}$ criterion
from the  distribution of $|y_{l}-y_{H}|$ with 10 fb$^{-1}$ of the integrated luminosity.
We perform the $\chi^{2}$ on this distribution because the signal and backgrounds shapes
are different and therefore could lead to stronger limits.
The $\chi^{2}$ criterion is defined as:
\begin{eqnarray}
\chi^{2}(\frac{\kappa_{u,c}}{\Lambda})=\sum_{i=bins}\frac{(s_{i}-b_{i})^{2}}{\Delta_{i}^{2}}
\end{eqnarray}
where $s_{i}$ denotes the number of signal events in the i-th bin of the 
$y_{l}-y_{H}$ distribution and $b_{i}$
is the number background events predicted by the standard
model in the i-th bin.
The $\chi^{2}$ criterion depends on anomalous couplings of $\kappa_{u,c}/\Lambda$.
In the $\chi^{2}$ definition, $\Delta_{i}=b_{i}\sqrt{\delta_{stat}^{2}+\delta_{syst}^{2}}$
where $\delta_{stat}$ is the statistical uncertainty and $\delta_{syst}$ denotes the term for considering
systematic uncertainties. Systematic uncertainties from the top quark mass, PDF, factorization and renormalization scales,
luminosity measurements and etc. are necessary for more realistic results. However, at this
level of analysis it is difficult to give estimations of all systematics. 
Therefore, a combined systematic uncertainty of $10\%$ is taken into account. 
The $68\%$ C.L. upper limits on the anomalous FCNC couplings are found to be:
\begin{eqnarray}
\frac{\kappa_{tug}}{\Lambda} \leq 0.014~{\text TeV}^{-1}~,~\frac{\kappa_{tcg}}{\Lambda} \leq 0.045~{\text TeV}^{-1}
\end{eqnarray}
Certainly, these limits could be improved using advanced methods
to separate signal from backgrounds  such as neural networks \cite{nn} and 
 boosted decision trees. The combination of the limits from this channel
with other channels also can lead to tighter bounds on the anomalous couplings.

In this analysis, we have not considered QCD multijet events. Because of its huge cross section,
a data-driven technique is needed to estimate the contribution of this background. However,
it is expected that the contribution of this background is negligible
after the requirement of one isolated lepton and the missing transverse energy.
Furthermore, requiring three b-tagged jets that two of them must have a mass in the
Higgs mass window is expected to suppress the QCD background.\\
The SM Single top plus Higgs, $tZj$, and $t\bar{t}Z$ events can also be sources of
backgrounds to our signal. The inclusive LO cross sections are
52 fb, 0.55 pb, and 1.02 pb, respectively. We have not included these backgrounds
in the analysis due to very small cross sections. After including the branching ratios
and applying the cuts, negligible number of events will be survived.\\
One of the main backgrounds to this analysis is $W+jets$.
The requirement of exactly three b-jets suppresses this background dramatically.
We expect that a full analysis with well developed algorithms for
b-tagging provides more precise and reliable results. Therefore, a full detector
analysis by the experimental collaborations is needed to confirm the results
that we obtained in this analysis.

\section{Simultaneous Probe of  $tqg$ and $tqH$ Couplings}

\begin{figure}
\centering
\includegraphics[width=8cm,height=7cm]{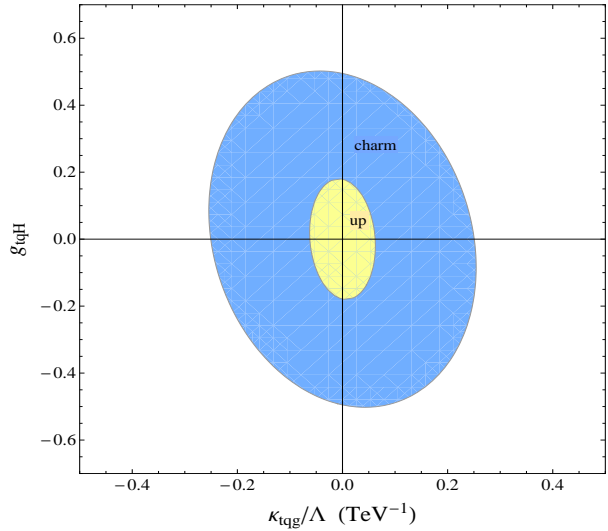}
\caption{The $3\sigma$ exclusion upper limits on the anomalous couplings $\frac{\kappa_{tqg}}{\Lambda}$
and $g_{tqH}$ for 10 fb$^{-1}$ of integrated luminosity at the LHC with the center-of-mass energy
of 14 TeV.}\label{contour}
\end{figure}

The final state of single top quark plus a Higgs boson can arise from 
both anomalous interactions $tqg$ and $tqH$. Both anomalous couplings 
come from dimension six operators. 
Therefore, in the presence of both couplings the anomalous single top quark in association with a Higgs boson production cross section 
can be parameterised as:
\begin{eqnarray}
\sigma(\frac{\kappa_{tqg}}{\Lambda},g_{tqH})[\text{pb}] = c_{tqg}\times (\frac{\kappa_{tqg}}{\Lambda})^{2}+c_{tqH}\times g_{tqH}^{2}+c_{int.}\times
\frac{\kappa_{tqg}}{\Lambda}\times g_{tqH}
\end{eqnarray}
where $\kappa_{tqg}/\Lambda$ is in TeV$^{-1}$ and $g_{tqH}$ is dimensionless.
The coefficients $c_{tqg}$, $c_{tqH}$, and $c_{int.}$ are determined with {\sc MadGraph}. After the 
preliminary cuts described in section 2, the coefficients are $c_{tu(c)g} = 5.6(1.05)$, $c_{tu(c)H}=0.09(0.01)$, and $c_{int.}=0.46(0.2)$.
The numbers in parantheses denotes the coefficients for the $tcg$ and $tcH$ couplings.
As it can be seen, the anomalous $tqg$ coupling can have larger contribution to the 
production of a single top quark in association with a Higgs boson.
After applying similar requirements to what explained in the previous section the $3\sigma$
exclusion limits on the anomalous $tqg$ and $tqH$ are extracted. Figure \ref{contour}
shows the $3\sigma$ exclusion regions in the plane of $(\kappa_{tqg}/\Lambda,g_{tqH})$
using 10 fb$^{-1}$ of the integrated luminosity in proton-proton collisions at 14 TeV.
In this plot, the smallest region shows the $3\sigma$ region for the anomalous 
interactions $tug$ and $tuH$ and the bigger one is the allowed region for $tcg$ and $tcH$. 
Because of the smaller contribution to the signal cross section looser bounds are obtained on
the $tqH$ couplings with respect to the $tqg$ couplings.

\section{Charge Ratio}

\begin{figure}
\centering
\includegraphics[width=7cm,height=6cm]{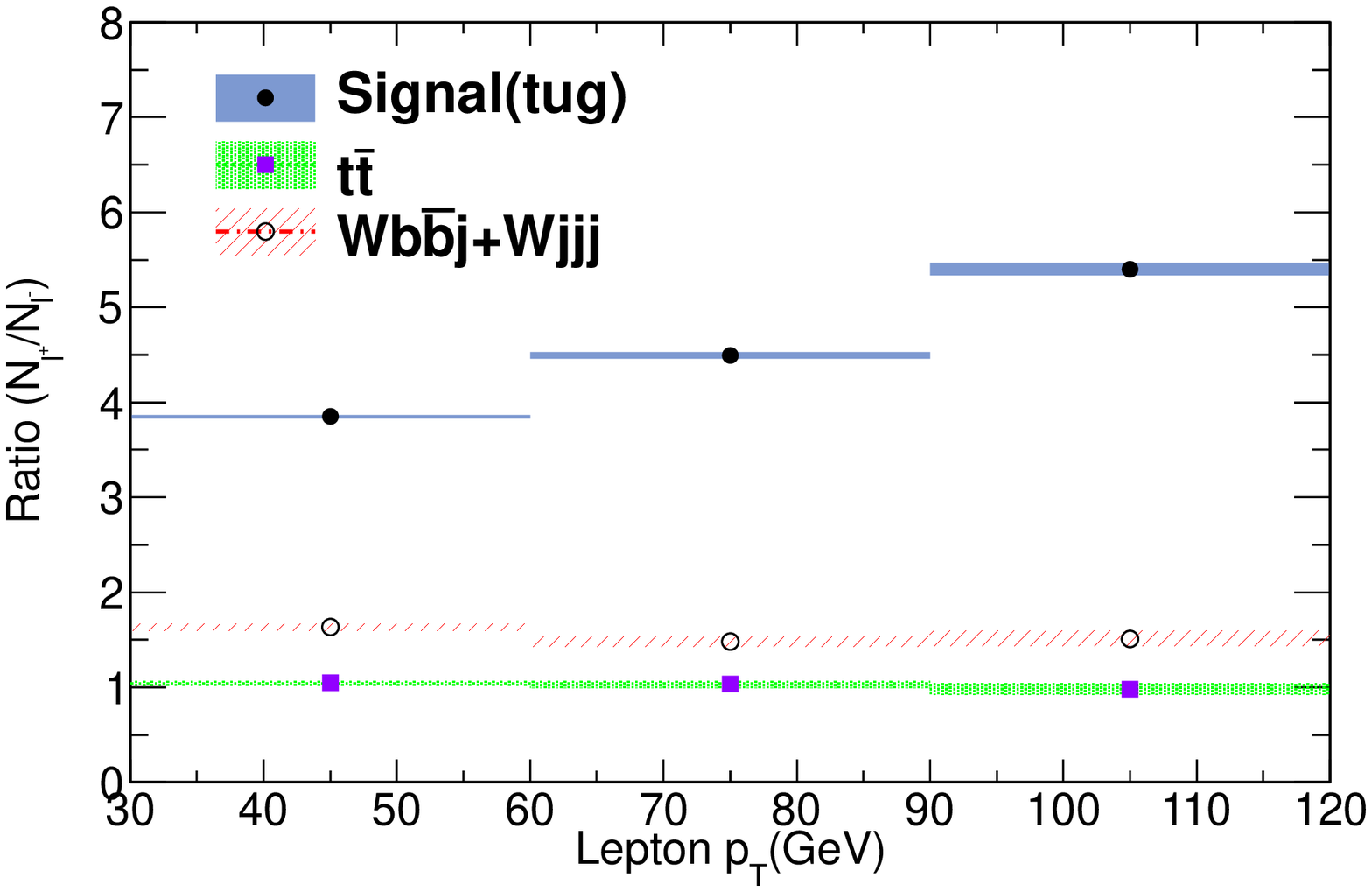}
\includegraphics[width=7cm,height=6cm]{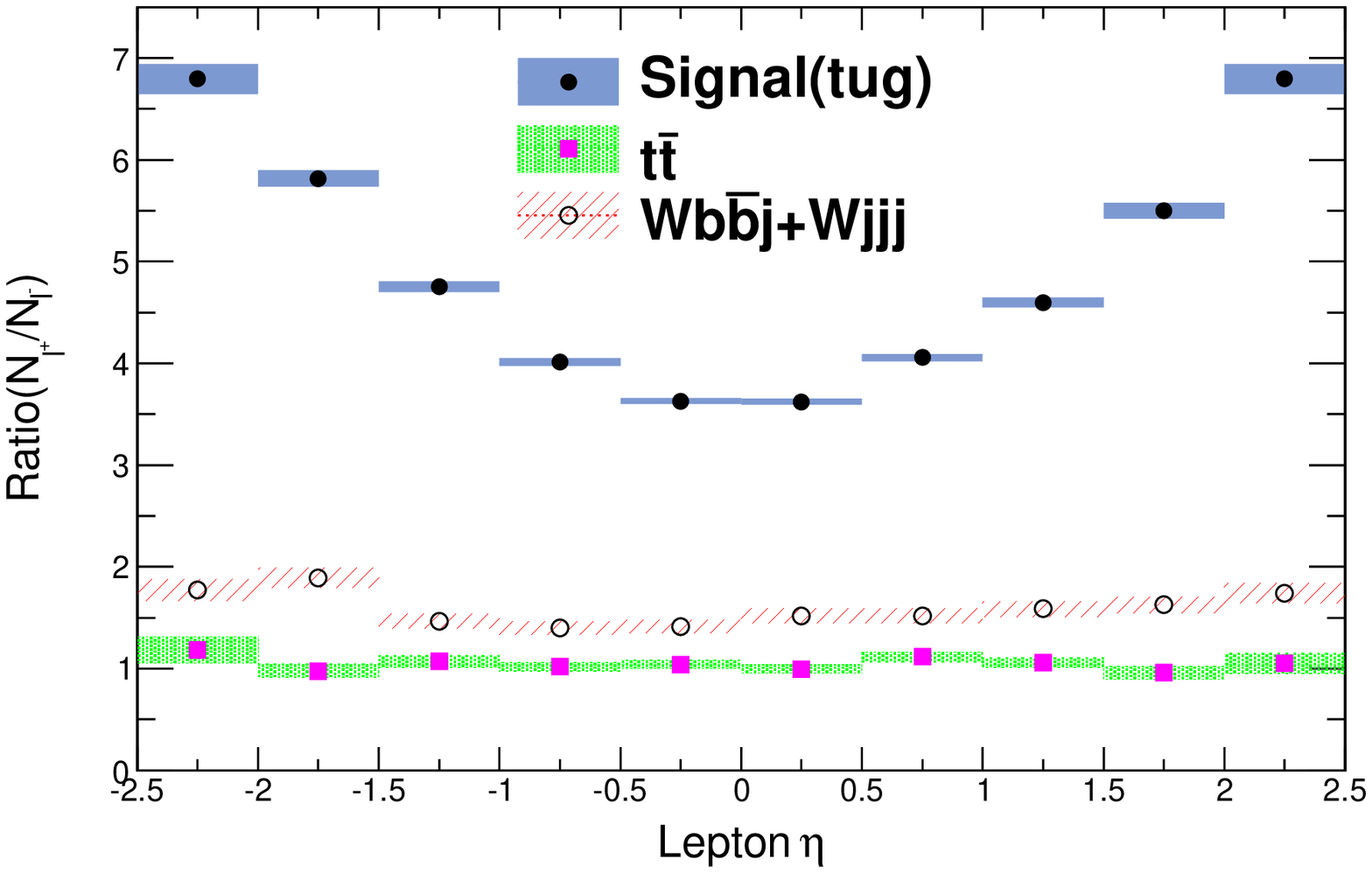}
\caption{The ratio of positive to negative leptons as a function of lepton $p_{T}$ (left) and lepton $\eta$ (right).
The uncertainty is only the statistical one.}
\label{ratio}
\end{figure}

One of the striking features of our signal, single top plus a Higgs boson
production at the LHC, is asymmetry between top and anti-top rates.  The cross section of top and anti-top
quarks are different at the LHC for the process of $g+u(\bar{u})\rightarrow t(\bar{t})+H$
because of the difference between the $u$-quark and $\bar{u}$-quark parton distribution functions
of proton. Since the $c$-quark and $\bar{c}$-quark parton distribution functions are similar, the
rates of top and anti-top quarks from the process of  $g+c(\bar{c})\rightarrow t(\bar{t})+H$
are expected to be similar.
In leptonic top decay, the top/antitop asymmetry is directly translated in a corresponding lepton charge
asymmetry. This is a reasonable assumption because the efficiencies of lepton selection and also fake charged
lepton contamination are almost independent of charge.
The dominant background to our signal is $t\bar{t}$ which is charge symmetric at leading order.
However, when the next-to-leading order corrections are included anti-top quarks prefer to be more central
than the top quarks. Therefore, more leptons will be observed than anti-leptons
in the central region of the detector.
The magnitude of this charge asymmetry is estimated to be around $1\%$ \cite{Kuhn:2011ri}.
QCD multi-jet background is expected to be perfectly charge-symmetric \cite{cms_st}. The only background which has charge
asymmetry among the main backgrounds is $W + jets$.
This nice feature of signal provides the possibility of reaching
the signal in the form of an excess in the ratio of positive to negative leptons after subtraction of the
expected contribution of the $W + jets$ background.
In such analysis one has to take into account the possibility of charge mis-measurement as well as any potential
differences in efficiency between the positive and negative leptons.
However, these are expected to be negligible in particular for muons.
In this analysis, we define a ratio $R$ as the number of events with positive charged lepton to
the number of events with negative charge. The inclusive values of $R$ for signal, $W+jets$ ($W+jjj$ and $Wb\bar{b}j$),
and $t\bar{t}$ are:
\begin{eqnarray}
R_{\text{signal}}  = 4.35 \pm 0.02,~R_{W+jets} = 1.57 \pm 0.03,~R_{t\bar{t}} = 1.04\pm 0.03
\end{eqnarray}
where the uncertainties are only statistical uncertainties.
As it can be seen the inclusive value of the charge ratio for signal is significantly larger than
the main backgrounds even around three time larger than the ratio of the charge asymmetric $W+jets$
background. It is important to note that the value of $R$ for signal is independent of the value of the
anomalous couplings $\kappa_{tug}/\Lambda$. Similar feature exists for direct top production
due to anomalous $tqg$ couplings which has been discussed in \cite{Gao:2011fx}.
In addition to the inclusive value of the charge ratio, we investigate the dependence of
the charge ratio $R$ for the signal and main backgrounds on the transverse momentum
and pseudorapidity of the charged lepton.
Figure \ref{ratio} shows the charge ratio $R$ as a function of lepton $p_{T}$ (left) and
lepton $\eta$ (right). As it can be seen, $R$ grows with increasing the lepton $p_{T}$
for signal while it is almost flat for $t\bar{t}$ and $W+jets$ backgrounds.
The charge ratio is around 3.8 for low $p_{T}$ leptons while it goes up to 5.4 for very
energetic charged leptons. This behavior can be understood by considering the
fact that the high $p_{T}$ lepton in the final state needs larger fraction of the
parton momentum from the proton PDF. It is well-known that the up quark PDF are much larger than
the anti-up quark PDF at large values of $x$ ($x$ is the fraction of the proton momentum which
a parton carries). Thus, at large lepton $p_{T}$, larger ratio is expected.

\begin{figure}
\centering
\includegraphics[width=7cm,height=6cm]{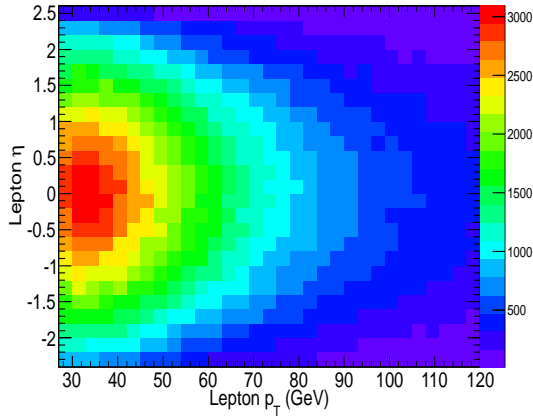}
\caption{The correlation between the transverse momentum and pseudorapidity of the charged lepton for signal events.}
\label{correlation}
\end{figure}

The ratio $R$ as a function of lepton $\eta$ is depicted in the right side of Figure \ref{ratio}.
Again for top pair events the ratio is almost flat and fluctuating around one while for $W+jets$
is very slowly increasing with $|\eta|$. For the signal, $R$ starts from 3.5 at $\eta \sim 0$ and grows
significantly up to 6.8 at $2.0 \leq |\eta| \leq 2.5$.
It is apparent that the ratio $R(p_{T})$ and $R(\eta)$ has a strong
discriminating power between signal and the main backgrounds. The
increasing behavior of the charge ratio with $|\eta|$ can be understood
by looking at Fig.\ref{correlation}. As it can be seen in this figure,
there is an apparent correlation between $p_{T}$ and $\eta$ of the charged lepton for the
signal events. Higher lepton $p_{T}$ events are correlated with larger lepton $\eta$.
Therefore, the large charge ratio for very energetic lepton would lead to the large
charge ratio in the forward/backward region. Indeed, there is a correlation between
$R(p_{T})$ and $R(\eta)$.

\begin{figure}
\centering
\includegraphics[width=7cm,height=6cm]{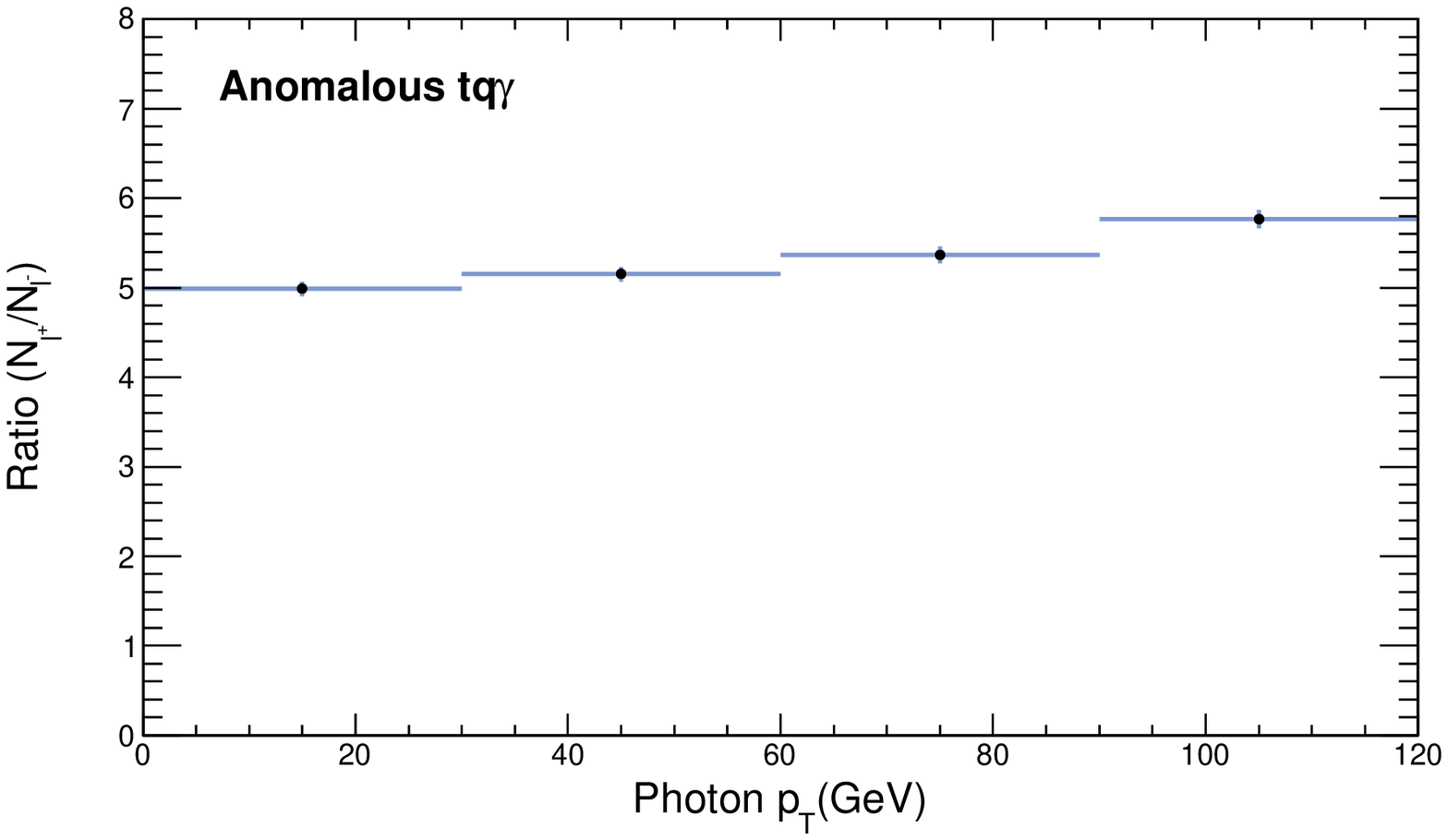}
\includegraphics[width=7cm,height=6cm]{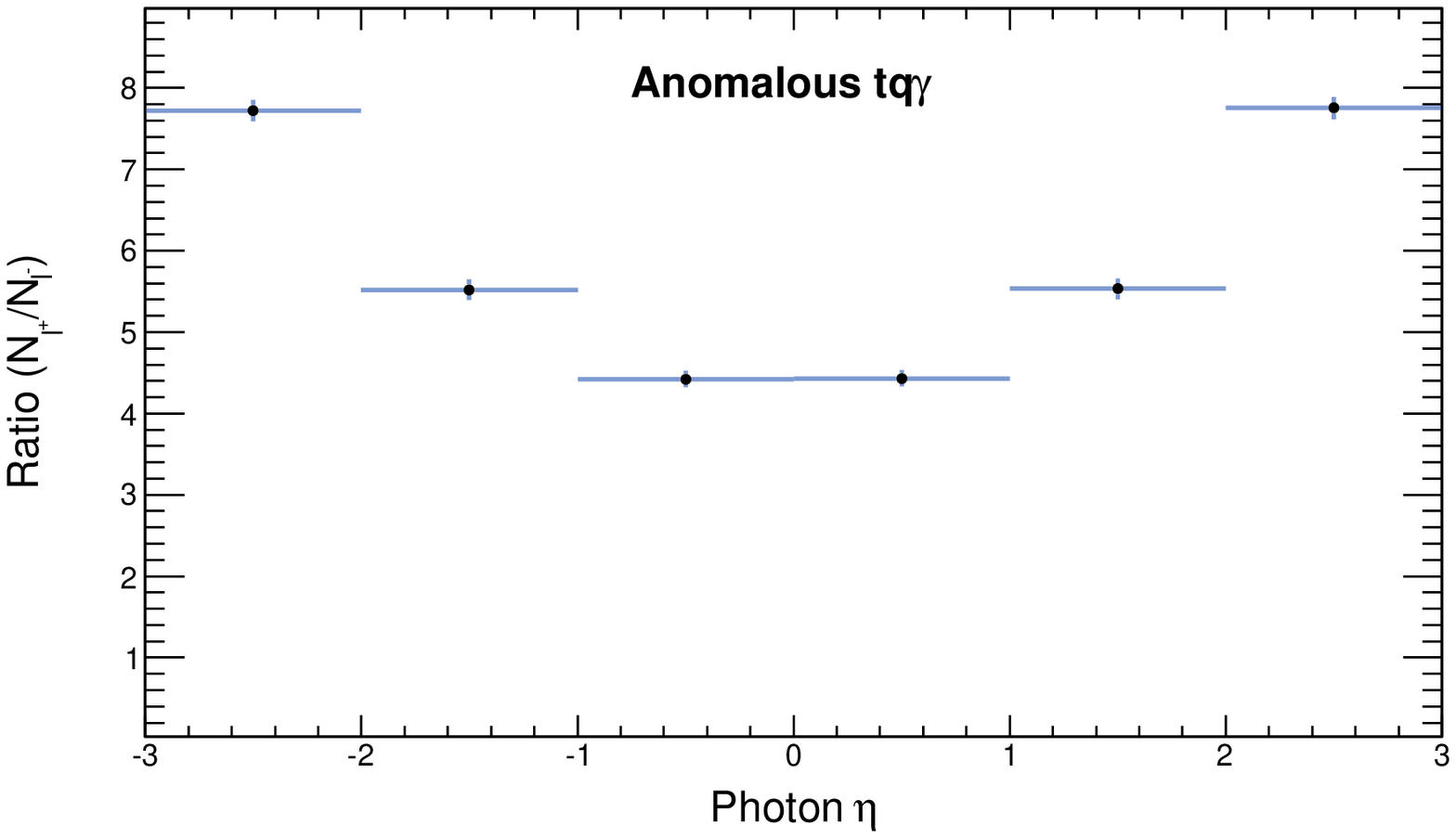}
\caption{The ratio of anti-lepton to lepton as a function of photon $p_{T}$ (left) and photon $\eta$ (right).}
\label{ratio11}
\end{figure}

It is important to note that the charge ratio is sensitive to the choice of parton distribution function (PDF) of the proton.
In \cite{cms_st}, the CMS Collaboration has measured the charge ratio in single top t-channel. The largest source of 
systematic uncertainty on the ratio is coming from the limited knowledge of proton PDF.
In this work, we have estimated the the uncertainties due to PDF by using the 44 members of the CTEQ6.6 PDFs. We have found
that the relative uncertainty due to PDF on the ratio $R$ is around $\Delta R/R = 7\%$.
The PDF uncertainty on the ratio $R$ varies in bins of the lepton $\eta$. For the central leptons
the PDF uncertainty is around $3\%$ which increases up to around $6-7\%$ for the leptons in the forward/backward region.
We also varied the factorizatio and renormalization scale to find uncertainty on the charge ratio. It is found to be less than $1\%$.\\
Apart from the ability of the charge ratio to discriminate between signal and backgrounds, upon the signal
discovery it can be used to determine that the signal comes from $t-u-g$ coupling or $t-c-g$ couplings.
Since the $t-c-g$ anomalous coupling has equal contribution in top and anti-top production the inclusive
and the differential charge ratio ($R(p_{T})$ and $R(\eta)$) have quite different values and behaviors
with the case that the signal originates from  $t-u-g$ anomalous coupling.\\
It is notable that similar charge ratio properties as mentioned in this section is applicable 
on the other channels of anomalous single top production in association with a vector boson or a Higgs boson.
Processes like $q+g \rightarrow t+\gamma$ (with anomalous interaction of $tq\gamma$ and $tqg$) and $q+g\rightarrow t+H$ 
(with anomalous couplings of $tqH$ and $tqg$) and also $q+g \rightarrow t+Z$ (with anomalous interaction of $tqZ$ and $tqg$) 
 \cite{Wang:2012gp},\cite{cmstz},\cite{Agram:2013koa}. As an example,
we show the charge ratio in the process of $q+g\rightarrow t+\gamma$ 
(with $tq\gamma$ anomalous interaction) as a
function of photon $p_{T}$ and $\eta$ in Fig.\ref{ratio11}.
An increasing behavior for the charge ratio
at large photon transvrse momentum and large rapidities can be seen.

\section{Conclusions}
In this work we propose to use the $pp\rightarrow t(\bar{t})+H$ process
to probe the anomalous $tug$ and $tcg$ couplings as a complementary
channel besides the other channels.
We concentrate on the leptonic decay of the top quark and the Higgs
boson decay to $b\bar{b}$ at the LHC with the center-of-mass energy of 14 TeV.
A set of kinematic variables have been proposed to discriminate between
the signal from backgrounds. After applying the selection, we show that
the LHC can probe the anomalous $tug(tcg)$ couplings down to 0.01 (0.04) TeV$^{-1}$
with 10 fb$^{-1}$ of integrated luminosity.
We also study the production of a signle top quark plus a Higgs boson 
coming from $tqg$ and $tqH$ anomalous couplings at the same time and derive 
the $3\sigma$ exclusion upper limits on the strengths of the anomalous couplings.
 We propose the charge ratio
versus transverse momentum and the pseudorapidity of the charge lepton
as a strong tool to discriminate between signal and backgrounds
as well as its ability to distinguish between the anomalous couplings
$tug$ and $tcg$. We have shown that in particular in the high$-p_{T}$
region or for the leptons in the forward/backward regions, the charge
ratio increases significantly. We have found that the charge ratio is robust 
against the variation of PDF and $Q$-scale.

\end{document}